\documentclass[aps,twocolumn,preprintnumbers,amsmath,amssymb]{revtex4}
\usepackage{dcolumn}
\usepackage{bm}
\usepackage{graphicx,subfigure,xcolor}
\usepackage{braket}
\usepackage[section]{placeins}

\newcommand{\ad}{a^{\dag}}
\newcommand{\bd}{b^{\dag}}
\newcommand{\Tr}{\hbox{Tr}}

\begin{document}

\title{Probing anharmonicity of a quantum oscillator in an optomechanical cavity}

\author{Ludovico Latmiral$^1$}
\email{ludovico.latmiral@hotmail.it}

\author{Federico Armata$^1$}
\email{f.armata@imperial.ac.uk}

\author{Marco G. Genoni$^2$}

\author{Igor Pikovski$^{3,4}$}

\author{M. S. Kim$^1$}

\affiliation{$^1$QOLS, Blackett Laboratory, Imperial College London, London SW7 2BW, United Kingdom}
\affiliation{$^2$Department of Physics and Astronomy, University College London, Gower Street, London WC1E 6BT, United Kingdom}
\affiliation{$^3$ITAMP, Harvard-Smithsonian Center for Astrophysics, Cambridge, MA 02138, USA}
\affiliation{$^4$Department of Physics, Harvard University, Cambridge, MA 02138, USA}

\begin{abstract}
We present a way of measuring with high precision the anharmonicity of a quantum oscillator coupled to an optical field via radiation pressure. Our protocol uses a sequence of pulsed interactions to perform a loop in the phase space of the mechanical oscillator, which is prepared in a thermal state. We show how the optical field acquires a phase depending on the anharmonicity. Remarkably, one only needs small initial cooling of the mechanical motion to probe even small anharmonicities.
Finally, by applying tools from quantum estimation theory, we calculate the ultimate bound on the estimation precision posed by quantum mechanics and compare it with the precision obtainable with feasible measurements such as homodyne and heterodyne detection on the cavity field. In particular we demonstrate that homodyne detection is nearly optimal in the limit of a large number of photons of the field, and we discuss the estimation precision of small anharmonicities in terms of its signal-to-noise ratio.
\end{abstract}

\maketitle

\section{\label{sec:level1}Introduction}

In the last years the field of quantum opto-mechanics has attracted significant interest, with the aim to control massive mechanical oscillators at the quantum level. In particular quantum optomechanical cavities \cite{aspelmeyer2014} have been investigated in great detail, and many research groups have proposed and  studied different implementations with moving end mirrors \cite{arcizet2006, gigan2006}, separate intra-cavity membranes \cite{thompson2008} or levitating nanospheres \cite{barker2010,chang2010,pflanzer2012} as mechanical oscillators.\\
Thanks to their peculiar properties, quantum optomechanical systems have been historically studied in the context of force sensing \cite{caves1980, braginsky1995} and have been recently proposed as a promising platform to test collapse models of quantum mechanics \cite{romeroisart2011, bahrami2014} and phenomenological models of quantum gravity \cite{pikovski2012, bawaj2015}. A major focus of research is now also devoted to the preparation of non-classical states of the mechanical motion, such as squeezed states \cite{kronwald2013, genoni2015a, genoni2015b,wollman2015, pirkkalainen2015}, single phonon excitations \cite{rips2012, qian2012, borkje2014} or even Schr\"odinger cat states \cite{bose1997, penrose2003, lombardo2015}.\\
In nearly every case cited above, the quantum mechanical oscillator is approximated harmonic, as the intrinsic anharmonic terms are considered small enough to be neglected. However it has been recently shown how the anharmonic/nonlinear regime can be accessed in different physical platforms. For instance, the effects of nonlinearities have been explored (and exploited) in mechanical resonators based on graphene and carbon nanotubes (see \cite{Dykman2012, Eichler2011} and references therein).
Also, in the case of a levitated nanosphere it has been shown that its thermal energy is sufficient to drive the motion of the oscillator into the nonlinear regime \cite{gieseler2013}; in another example, electrostatic gradient forces are exploited in order to enhance the intrinsic quartic anharmonicity of a nanomechanical resonator \cite{rips2014}. Finally, the non-linear dynamics and the cooling of a levitating nanosphere motion in a hybrid electro-optical trap has been experimentally achieved in Ref.\cite{fonseca2015}.\\
Aside from perturbing the behavior and results that one would obtain in the harmonic case, anharmonicity gives rise to new interesting quantum peculiarities. For instance, Milburn and Holmes studied the quantum and classical dynamics of an anharmonic oscillator in phase space showing that a decoherence reduction results in quantum-to-classical-transition \cite{milburn1986}. On the other hand, anharmonicity has been proven to be a resource to generate non-classical quantum states \cite{joshi2011, lu2015, teklu2015}, and a measure able to quantify the non-linearity of a quantum oscillator has been recently proposed \cite{paris2014}. \\
Given these premises, it is now desirable to design a protocol able to measure anharmonicity, in order to efficiently analyze its contribution to the dynamics and its effect on the experimental results.

In this work we present a scheme to estimate the anharmonicity of a quantum mechanical oscillator in an optomechanical cavity. Specifically, we provide a method based on the measurement of the phase shift of an optical field after its interactions with a quantum anharmonic oscillator. High precision can be achieved requiring a feasible initial cooling of the oscillator and the protocol reveals to be robust against losses. Furthermore, we give the ultimate quantum bound on the precision achievable through this setup, comparing it to the one obtainable with standard measurements on the optical field, such as homodyne and heterodyne measurements.
The manuscript is structured as follows: in Sec. \ref{sec:The model} we introduce the model of an optomechanical cavity and we present a \textit{pulsed} scheme that has been already studied in literature to measure the quantum dynamics of opto-mechanical systems \cite{pikovski2012, braginsky1995, vanner2011, khosla2013}. In particular we show how the unitary operator that describes the overall evolution of the system is related to a displacement operation that drives the mechanical oscillator along a closed path in phase space. Sec. \ref{sec:Anharmonic Displacement} is dedicated to the computation of the anharmonic contribution to the evolution and its effect on the phase shift acquired by the optical field. In Sec. \ref{sec:Estimation} we apply tools from quantum estimation theory and calculate the Quantum Fisher Information (QFI) and the Fisher Information (FI) for different measurement schemes to quantify how performing our estimation method is. We finally evaluate the corresponding signal-to-noise ratio to better discuss the estimability of small values of anharmonicity against noise fluctuations. Sec. \ref{sec:Conclusion} is devoted to our conclusive remarks.

\section{\label{sec:The model}The model}
We consider a single mode optical field of frequency $\omega_c$ coupled to a quantum anharmonic oscillator of mass $m$ and frequency $\omega_m$ in an optomechanical cavity of length $L$. The effective Hamiltonian describing our system in a frame rotating at the laser frequency on resonance with the optical cavity frequency is $\mathcal{H}=\mathcal{H}_0+\mathcal{H}_{int}$, where
\begin{equation}\label{H_0}
\mathcal{H}_0=\frac{1}{2}\hbar\omega_m\left(X_m^2+P_m^2\right)+\frac{\gamma}{4}\hbar\omega_mX_m^4
\end{equation}
is the free Hamiltonian of the mechanical oscillator, with:  $X_m=(\bd_0+b_0)/\sqrt{2}$ and $P_m=i(\bd_0-b_0)/\sqrt{2}$ its quadratures operators and $\gamma\ll 1$ the quartic anharmonic parameter. We will replicate in Appendix B all the results for the case of a cubic anharmonicity ($\delta/3\hbar\omega_m X^3_m$). The interaction Hamiltonian is given by \cite{law1995}
\begin{equation}\label{H_int}
\mathcal{H}_{int}=\hbar g n_cX_m
\end{equation}
where $n_c=\ad a$ is the photon number operator for the cavity field and $g=(\omega_c/L)\sqrt{\hbar/m\omega_m}$ is the coupling strength. In the case of a pulsed regime the interaction is much faster than a mechanical period and the mechanical position is essentially constant during the interaction. We can then neglect the free evolution of the harmonic oscillator during the interaction time and the dynamics can be described by the unitary operator \cite{vanner2011}
\begin{equation}\label{U}\begin{split}
  U=e^{i\lambda n_cX_m}
\end{split}\end{equation}
with $\lambda=g/k$ the rescaled coupling constant and $k$ the cavity decay rate which, in the pulsed regime, satisfies the bad cavity limit $k\gg\omega_m$.\\
Loosely speaking, the operator in Eq. \eqref{U} can be pictured as a displacement operation by $\lambda n_c/\sqrt{2}$ along $P_m$ in the oscillator phase space (the sentence is rigorous if the cavity field is prepared in a Fock state $|n\rangle$).
\begin{figure}[ht!]
\centering
\includegraphics[scale=0.24]{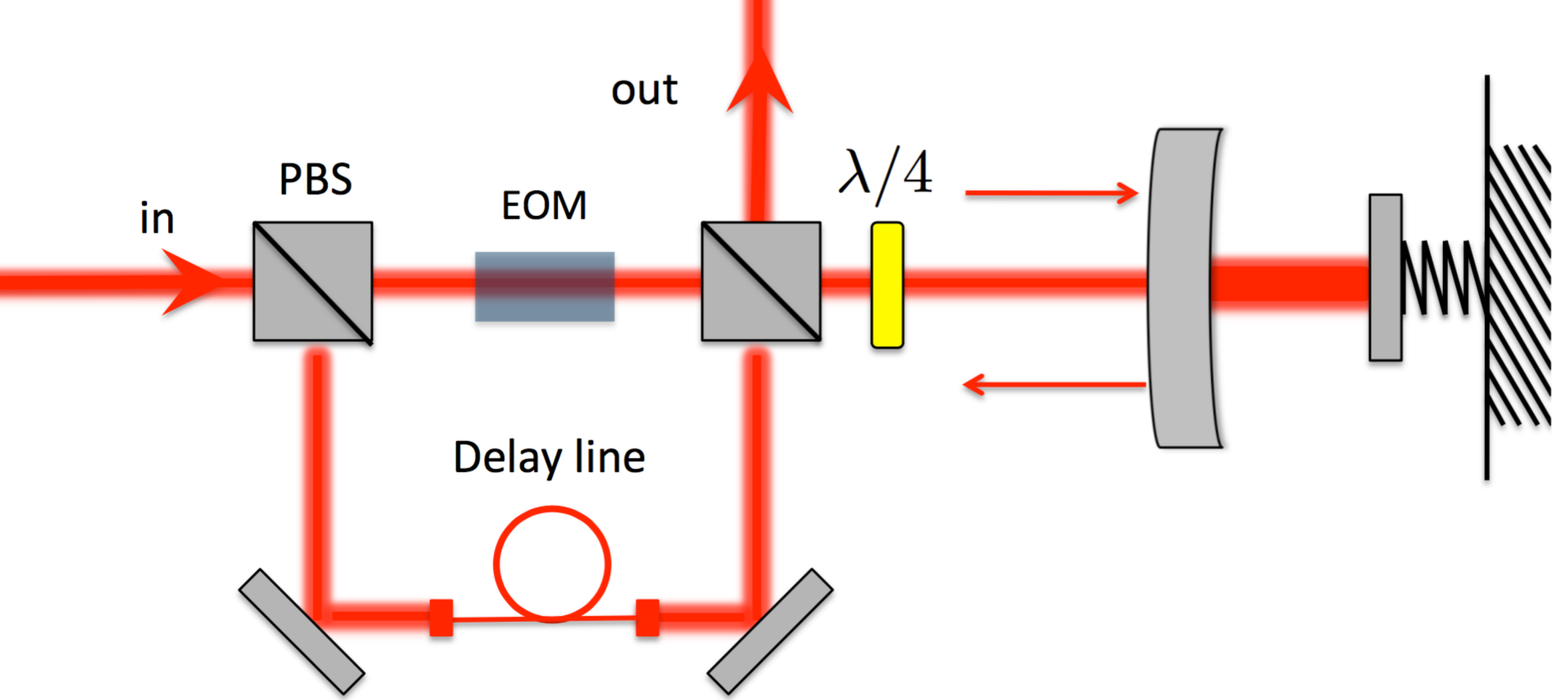}
\caption{Schematic representation of the model. The laser pulse enters into an optomechanical cavity and escapes entering in a delay loop for an engineered time. The apparatus composed by the polarizing beam splitters (PBSs), the $\mathcal{\lambda}/4$ wave plate and the Electro-optic Modulator (EOM) is used to rotate the polarization before and after each pulse. After the last interaction the EOM does not rotate the polarization and the pulse escapes the cavity, being measured interferometrically with respect to a reference field.
\label{pulsedscheme}}
\end{figure}
As soon as the interaction vanishes, the oscillator is free to evolve under the Hamiltonian $\mathcal{H}_0$ and $X_m$ and $P_m$ start to interchange themselves accordingly. We can therefore drive the oscillator along closed loops in phase space by selecting the appropriate  time between consecutive pulsed interactions. More specifically, we imagine that the same light pulse enters the cavity, escapes after a short interaction (lasting a time $1/k$) and waits in an engineered loop before being injected again (see Fig.\ref{pulsedscheme}).
\section{\label{sec:Anharmonic Displacement} The estimation protocol}

Using four pulsed interactions, each described by the operator in Eq. \eqref{U} with a free mechanical evolution in between, we drive the quantum oscillator along a loop in its phase space. The total evolution operator can be written as
\begin{equation}\label{Displacement4}\begin{split}
 U=e^{i\lambda n_c X_m(\frac{3\tau}{4})}e^{i\lambda n_c X_m(\frac{\tau}{2})}e^{i\lambda n_cX_m(\frac{\tau}{4})}e^{i\lambda n_c X_m}
\end{split}\end{equation}
where $\tau=2\pi/\omega$ is the mechanical period of the quantum anharmonic oscillator.
To explicitly compute Eq.\eqref{Displacement4} we need first to solve the dynamics of a quantum anharmonic oscillator. Evolution of quadrature operators can be obtained from Heisenberg evolution for annihilation (creation) operator \cite{landau1976,manko1982}, which reads (at the first order in $\gamma$)
\begin{widetext}
\begin{equation}\label{b-time}\begin{split}
b(t) \simeq b_0e^{-i\omega t}+\frac{\gamma}{4}\left[\left(e^{-i\omega t}-e^{+3i\omega t}\right)\frac{b_0^{\dag 3}}{4}+\left(e^{-3i\omega t}-e^{-i\omega t}\right)\frac{b_0^{3}}{2}+\left(e^{-i\omega t}-e^{i\omega t}\right)\frac{3}{2}\bd_0\left(1+b_0^{\dag}b_0\right)\right]
\end{split}\end{equation}
\end{widetext}
with
\begin{equation}\label{omega}\begin{split}
  \omega=\omega_m+\frac{3}{8}\gamma\omega_m\; (2+|A|^2),
\end{split}\end{equation}
$|A|$ being the oscillation amplitude for the unperturbed harmonic oscillator. We point out that to be consistent with the perturbation approach we need the additional requirement $\gamma |A|^2=\gamma (\lambda N_p)^2\ll 1$, with $N_p=\langle n_c\rangle$ the average number of photons of the cavity field. By using Eq. \eqref{b-time} we find the quadrature operators at times $t=0,\tau/4,\tau/2,3\tau/4$,
\begin{equation}\label{Xmoretime}\begin{split}
  X_m(0)&=X_m\\
  X_m\left(\frac{\tau}{4}\right)&\simeq P_m+i\frac{\gamma}{4\sqrt{2}}\Delta\\
  X_m\left(\frac{\tau}{2}\right)&\simeq -X_m\\
  X_m\left(\frac{3\tau}{4}\right)&\simeq -P_m-i\frac{\gamma}{4\sqrt{2}}\Delta
\end{split}\end{equation}
with $\Delta=b_0^3-b^{\dag 3}_0-3b_0^{\dag}+3b_0-3b_0^{\dag 2}b_0+3b_0^{\dag}b_0^2$ the deformation due to the anharmonic evolution. \textit{We remark that at time $t =\tau$ the oscillator returns to its initial position (at the first order in $\gamma$).} As we are going to discuss, this is an essential requirement, since only for closed loops field and oscillator can become uncorrelated after a sequence of interactions \cite{mancini1997, bose1997}. In particular, to close the loop we need the anharmonic frequency (see Eq.\eqref{omega}) that actually is a function of the anharmonic parameter we want to estimate. This is a common situation in local quantum estimation theory and can be worked out by  subsequent adaptive measurements  \cite{higgins2007,brivio2010,berni2015}. Moreover, since our final goal is to measure the anharmonic parameter via an interferometric scheme (e.g. homodyne and heterodyne detection), we can ensure the closure of the loop also by looking at the visibility of the interference fringes \cite{penrose2003, armata-latmiral}. However, we should also remark that this is not the case for the cubic anharmonicity (see Appendix B) that does not alter the mechanical frequency. This peculiarity can be exploited to distinguish the two anharmonicities by only looking at the inteferometric pattern.

We are now interested in the reduced dynamics of the cavity field, {\em i.e.} we want to compute the completely-positive map $\mathcal{E}$ defined as
\begin{align}
\mathcal{E}(\varrho_0) &= \Tr_m [ U \varrho_0 \otimes \nu U^{\dag} ]
\end{align}
where $\Tr_m[\bullet]$ denotes the partial trace on the mechanical oscillator, while $\varrho_0$ and $\nu$ denote respectively the initial state of the cavity field and of the mechanical oscillator. In the following we will focus on the case where the oscillator is initially prepared in a state diagonal in the Fock basis, {\em i.e.} $\nu = \sum_n \nu_n |n\rangle\langle n|$, that comprises Gibbs thermal state. In order to obtain this map, we can calculate the evolution operator in Eq. \eqref{Displacement4} by substituting the expressions for the quadrature operators in Eq. \eqref{Xmoretime} (see Appendix A for further details). After some algebra, given the assumptions described above, the evolution of the optical field after a closed loop reads (at the first order in $\gamma$ and in the limit $\lambda^2\langle n_c\rangle^2\gg \bar{n}$)
\begin{align}
\label{Displacement-final}
\mathcal{E}(\varrho_0) &\simeq \xi_{\sf eff} \varrho_0 \xi_{\sf eff}^\dag \:, \\
\textrm{with}& \:\:\:\; \xi_{\sf eff} = \exp \{i ( \lambda^2 n_c^2 - \frac{\gamma}{2}(\lambda^4 n_c^4+3\lambda^2 n_c^2 \bar{n}))\} \:,
\end{align}
where $\bar{n}$ is the average thermal phonon number.
We thus obtain an effective unitary operator $\xi_{\sf eff}$ acting on the cavity field, retaining all the information on the dynamics, and in particular on the anharmonicity parameter $\gamma$. We notice that the field experiences a Kerr-nonlinearity when it enters into the optomechanical cavity \cite{aldana2013}. Also, we remark that Eq. (\ref{Displacement-final})  is valid for any initial state of the oscillator diagonal in the Fock basis, such as Gibbs thermal state. This is one of the main results of this paper, as the estimation of the anharmonicity $\gamma$ relies on doable cooling of the mechanical oscillator. Indeed, the mild condition on the average number of thermal phonons $\lambda^2\langle n_c\rangle^2\gg \bar{n}$ guarantees that after a period the oscillator is uncorrelated to the field and closes the loop in phase space.

\noindent
Since our protocol relies on having the same light pulse for each interaction, it is worth estimating losses that might occur in the delaying fiber loops. The ratio between consecutive pulses can be modeled as $\lambda_{i+1}/\lambda_{i}=1-\epsilon$. The intensities of the four pulsed interactions in Eq. \eqref{Displacement4} and the resulting effective map (acting only on the cavity field) will be accordingly modified. Specifically, losses will affect the evolution operator, and as a consequence, mirror and field will be correlated after a loop. The overall noise on the anharmonic evolution can be neglected when $\epsilon\bar{n}\ll \langle n_c\rangle$, which is commonly satisfied in todays experiments (more details on this model are reported in Appendix C).

Supposing that all the previous conditions are satisfied, we can calculate the mean value of the optical field after a four-pulse interaction. If the cavity field is initially prepared in a coherent state $\varrho_0=\ket\alpha\bra\alpha$, the phase reads (in the limit $\gamma\lambda^4N_p^3\ll 1$ and $\lambda^2N_p^2\gg \bar{n}$)
\begin{equation}\label{afield}\begin{split}
  \langle a\rangle \simeq \langle\alpha|\xi_{\sf eff}^{\dag}a\xi_{\sf eff}|\alpha\rangle\simeq\alpha\langle a\rangle_0e^{-i\frac{\gamma}{2}\lambda^4(4N_{p}^3+18N_p^2+10N_p+1)}
\end{split}\end{equation}
with $\langle a\rangle_0=e^{i\lambda^2-N_{p}(1-e^{i2\lambda^2})}$ the phase acquired by the field for a harmonic dynamics, and where now $N_p = \langle n_c \rangle = |\alpha|^2$. As can be seen in Eq. \eqref{afield}, after a loop of the oscillator, the phase shift acquired by the optical field is independent of mechanical initial states, though it retains all the information on the dynamics.\\

\section{\label{sec:Estimation} Estimation properties of the anharmonic parameter}

\begin{figure}[h!]
\centering
\includegraphics[scale=0.38]{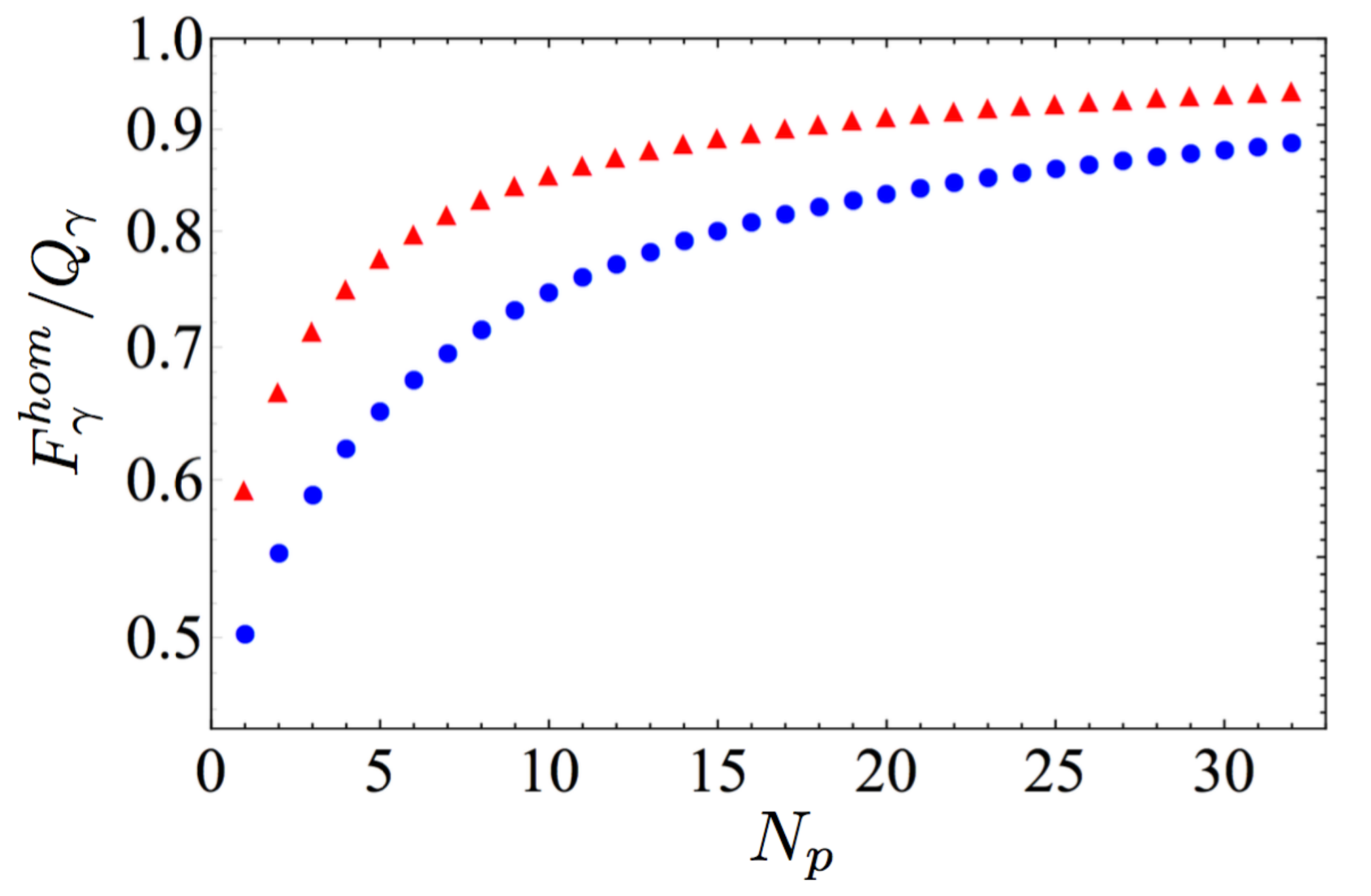}
\caption{Ratio $F^{\sf hom}_\gamma/Q_\gamma$ for cubic (red triangles) and quartic (blue dots) anharmonicities as functions of the average number of photons $N_p$. Experimental parameters are set as  $\lambda\sim 1.5\times 10^{-5}$, $\gamma =10^{-25}$ and the phase $\phi$ is optimized to $\phi=\pi/2$.
\label{homodyne_photon}}
\end{figure}

In order to assess how well one can estimate the anharmonicity parameter $\gamma$ through our measurement scheme, we are going to exploit tools from local quantum estimation theory \cite{paris2009}, deriving the ultimate bounds on the estimation precision and comparing them with the bounds corresponding to practical measurement schemes.
We start by calculating the QFI corresponding to the parameter $\gamma$ for the output state (\ref{Displacement-final}), under the assumptions previously discussed. As the effective dynamics is unitary, for an initial pure coherent state $|\alpha\rangle$, the output state will still be pure, {\em i.e.} $|\psi_\gamma\rangle = \xi_{\sf eff} |\alpha\rangle$, and the QFI can be evaluated as follows
\begin{equation}\label{QF}\begin{split}
Q_\gamma &=4\left(\langle\psi'_{\gamma}|\psi'_{\gamma}\rangle-|\langle\psi'_{\gamma}|\psi_{\gamma}\rangle|^2\right)\\
  &=\lambda^8\left(\langle\psi_{\gamma}|n_c^{8}|\psi_{\gamma}\rangle-\langle\psi_{\gamma}|n_c^4|\psi_{\gamma}\rangle^2\right)\\
  &= 16 \lambda^8 N_p^7 + O(N_p^6),
\end{split}\end{equation}
where $|\psi'_{\gamma}\rangle$ is the derivative of the state with respect to the anharmonic parameter. The QFI sets the ultimate lower bound on the estimation precision for the parameter $\gamma$ (quantified by the variance of an unbiased estimator), through the so-called quantum Cram\'er-Rao theorem that reads
\begin{equation}\label{CRB-t}\begin{split}
{\rm Var}(\gamma)\geq\frac{1}{M Q_\gamma}\gtrsim \frac{1}{16 M \lambda^8N_{p}^7} \:,
\end{split}\end{equation}
where $M$ denotes the number of measurements performed. We deduce from Eq.\eqref{CRB-t} that the estimation is highly enhanced by the Kerr-nonlinearity in Eq.\eqref{Displacement-final}, where the anharmonic contribution scales as $\sim\gamma\lambda^4n_c^4$.
The quantum bound is always in principle achievable for a single parameter, in the sense that there exists a POVM whose (classical) FI is equal to the QFI. To evaluate if feasible measurements are optimal we proceed by calculating the corresponding FI which in general reads
\begin{equation}\label{FI}\begin{split}
F_\gamma=\int d\bullet\frac{(\partial_{\gamma}p(\bullet|\gamma))^2}{p(\bullet|\gamma)} \:,
\end{split}\end{equation}
where $p(\bullet|\gamma)$ is a generic conditional probability of obtaining the measurement outcome $\bullet$, given the value of the parameter $\gamma$. In the following we will focus on two measurement strategies for the cavity field: homodyne and heterodyne detection.\\
Homodyne detection corresponds to a projection on quadrature operators eigenstates, $X_{\phi}|x\rangle_{\phi}=x|x\rangle_{\phi}$, where $X_{\phi}=x_c \cos\phi+p_c\sin\phi$, and the pair of operators $(x_c, p_c)$ denote respectively the {\em position} and {\em momentum} operators for the cavity field. In the Fock basis, we can write quadrature operator eigenstates as \cite{ferraro2005}
\begin{equation}\label{xphi}\begin{split}
  |x\rangle_{\phi}=e^{-x^2/2}\left(\frac{1}{\pi}\right)^{1/4}\sum_{m=0}^{\infty}\frac{H_m(x)}{2^{m/2}\sqrt{m!}}e^{-im\phi}|m\rangle,
\end{split}\end{equation}
where $H_m(x)$ is the $m$-th Hermite polynomials. The conditional probability of obtaining the outcome $x$, given $\gamma$, is
\begin{equation}\label{p-x}\begin{split}
  p(x|\gamma)&=|{}_\phi\langle x|\psi_{\gamma}\rangle|^2\\
  &=\frac{e^{-(|\alpha|^2+x)}}{\sqrt{\pi}}\left|\sum_{m=0}^{\infty}\frac{\alpha^mH_m(x)}{2^{\frac{m}{2}}m!}e^{im\left[\phi-\lambda^2m(1+\frac{\gamma}{2}\lambda^2m^2)\right]}\right|^2.
\end{split}\end{equation}
Unfortunately, there is no analytical way to compute this series, however, by fixing all the parameters \{$\alpha,\;\phi,\;\lambda,\;\gamma$\} and by varying the measurement outcome $x$, we can numerically evaluate the integral and find the FI as in Eq. (\ref{FI}). We show in Fig.\ref{homodyne_photon} the ratio between the homodyne FI and the corresponding QFI by optimizing the phase $\phi$.  As it can be seen from Fig. \ref{homodyne_photon} the larger the photon number is the more the ratio $F^{\sf hom}_\gamma / Q_{\gamma}$ approaches one. We also observe that we already reach a very good agreement with $30$ photons though this is actually very low compared to the number of photons in a standard optomechanical cavity setup. This clearly shows that homodyne detection is an advantageous method to probe anharmonicity with arbitrarily high precision, safely conjecturing its optimality in the limit of large number of photons.\\

\noindent
On the other hand, heterodyne detection corresponds to a projection on a coherent state $|\eta\rangle$, which can be performed through a double-homodyne detection scheme \cite{genoni2014}. The corresponding conditional probability is given by
\begin{equation}\label{p-eta}\begin{split}
p(\eta|\gamma)&= |\langle \eta | \psi_\gamma\rangle |^2  \\
&=e^{-(|\alpha|^2+|\eta|^2)}\left|\sum_{m=0}^{\infty}\frac{\alpha^m\eta^{*m}}{m!^2}e^{-i\lambda^2m^2(1+\frac{\gamma}{2}\lambda^2m^2)}\right|^2.
\end{split}\end{equation}
The FI can be computed by integration over in the whole complex plane spanned by coherent states
\begin{equation}
F^{\sf het}_\gamma=\frac{1}{\pi}\int d^2\eta \;\; \frac{(\partial_{\gamma}p(\eta|\gamma))^2}{p(\eta|\gamma)},
\end{equation}
where the dependence on the phase parameter has dropped out, as opposed to the case of homodyne detection. Also in this case the FI has been evaluated numerically for an initial coherent state with up to $35$ photons. Our numerical results show that the optimality of heterodyne measurement, quantified by the ratio between FI and QFI, does not depend on any parameter, being the ratio fixed to $F^{\sf het}_\gamma/Q_\gamma = 0.5$. We thus  conclude that it is much more convenient to perform a homodyne measurement on the cavity field, in order to estimate the anharmonicity with higher precision, and nearly quantum limited.

As we are dealing with very small values of the parameter to be estimated, the \emph{signal-to-noise ratio} is an important figure of merit that has to be considered. It tells us how the effective contribution of the physical quantity we want to measure compares to the noise. More specifically, bearing in mind Cram\'er-Rao bound theorem, we can define for any parameter $\zeta$ its signal-to-noise ratio $R_\zeta$ and derive the upper bound:
\begin{equation}
R_\zeta=\frac{\zeta^2}{{\rm Var}(\zeta)}\leq \zeta^2 M Q_\zeta \:,
\end{equation}
where $Q_\zeta$ denotes the QFI for the parameter of interest and $M$ is the number of measurements performed. An essential requirement for efficient metrology is to achieve a significant value of the signal to noise ratio $R_\zeta > 1$ with a reasonable number of experimental runs. \\
In our specific case, in the limit of large number of photons we get
\begin{equation}
\label{signaltonoise_system}
\begin{split}
R_\gamma^{(4)} \lesssim& \:16\gamma^2\lambda^{8}N_p^{7}M \: \\
R_\gamma^{(3)} \lesssim&\: \frac{16}{9}\delta^2\lambda^{6}N_p^{5}M \:,
\end{split}
\end{equation}
for a quartic and a cubic anharmonicity, respectively, and where we have shown before that these bounds may be in principle achievable via homodyne detection in the limit of large number of phonons. \\
 If we substitute the usual values of cavity parameters in Eq.\eqref{signaltonoise_system}, {\em e.g.} $N_p \sim 10^9$ and $\lambda \sim 10^{-4}$, and consider $M \sim 10^4$ number of experimental runs (which still allows us to use optimal asymptotic estimators, such as the Bayesian or the MaxLik estimator), our results show that one can in principle probe anharmonicities as low as $\gamma \sim 10^{-20}$ for the quartic case and $\delta \sim 10^{-15}$ for the cubic case. Eventually, we observe that for these values of the parameters all the assumptions that we have made (i.e. $\gamma\lambda^4N_p^3\ll 1$, $\lambda^2N_p^2\gg\bar{n}$, $\epsilon\bar{n}\ll N_p$) are satisfied for temperatures of a few kelvins, which can be easily achieved through dilution refrigeration.
\section{\label{sec:Conclusion}Conclusions}
We have presented a protocol to estimate the anharmonicity of a mechanical oscillator relying on a four-pulse interaction with an optical field. Under reasonable initial cooling the output oscillator and optical field states are uncorrelated; specifically, the oscillator returns to its initial position, while the cavity field undergoes an effective unitary operator which retains information on the anharmonicity of the mechanics. Since a frequency shift is only obtained in the case of quartic anharmonicities, and not in the cubic case, the scheme can also discriminate between the two. By using tools from local quantum estimation theory, we have also derived the ultimate bounds on the estimation precision, showing how this can be arbitrarily high by increasing the number of photons of the initial coherent state. Furthermore, we have shown the performances of homodyne detection, conjecturing its near-to-optimality in the limit of large number of photons. Finally, we have shown the efficiency of our method in estimating small anharmonicities by considering state-of-the-art values of the optomechanical parameters.

\section*{ACKNOWLEDGMENTS}
The authors wish to thank Tommaso Tufarelli for useful discussions on the subject of this paper. MSK acknowledges support from EPSRC through EP/J014664/1. FA and MSK acknowledge the financial support from the Marie Curie Project no. PITN-GA-2012-317232. MGG acknowledges support from EPSRC through grant EP/K026267/1. IP acknowledges support by the NSF through a grant to ITAMP.\\

\appendix
\section{\label{appendix1}Anharmonic displacement operator and Phase}

In this section, we sum up the main steps that lead to Eq. \eqref{Displacement-final}. By substituting Eqs. \eqref{Xmoretime} in \eqref{Displacement4} we get
\begin{equation}\label{D-x4-1}
U \simeq e^{i\lambda n_c(-P_m-i\frac{\gamma}{4\sqrt{2}}\Delta)}e^{-i\lambda n_c  X_m}e^{i\lambda n_c(P_m+i\frac{\gamma}{4\sqrt{2}}\Delta)}e^{i\lambda n_c X_m}.
\end{equation}
If now we apply Zassenhaus formula \cite{suzuki1977}, the first and third terms can be rewritten, respectively, as (to the first order in $\gamma$)
\begin{equation}\label{D-x4-2}\begin{split}
e^{i\lambda n_c(-P_m-i\frac{\gamma}{4\sqrt{2}}\Delta)}\simeq&e^{-i\lambda n_c  P_m}e^{\frac{\gamma}{4\sqrt{2}}f_1(b_0,b^{\dag}_0)}\\
e^{i\lambda n_c(P_m+i\frac{\gamma}{4\sqrt{2}}\Delta)}\simeq&e^{i\lambda n_c  P_m}e^{\frac{\gamma}{4\sqrt{2}}f_2(b_0,b^{\dag}_0)}
\end{split}\end{equation}
where
\begin{equation}\label{f}\begin{split}
&f_1(b_0,b^{\dag}_0)=\lambda n_c\Delta-\frac{3}{\sqrt{2}}\lambda^2 n_c^2(b^{\dag 2}-b^{2})+\sqrt{2}i\lambda^3 n_c^3P_m\\
&f_2(b_0,b^{\dag}_0)=-\lambda n_c\Delta-\frac{3}{\sqrt{2}}\lambda^2 n_c^2(b^{\dag 2}-b^{2})-\sqrt{2}i\lambda^3 n_c^3P_m.
\end{split}\end{equation}
Switching the latter factors in Eq. \eqref{D-x4-2} to the left and right respectively by iteratively applying Zassenhaus expansion we obtain the evolution operator at the first order in $\gamma$
\begin{equation}\begin{split}
U\simeq&\left(1+\frac{\gamma}{4\sqrt{2}} F_1(b_0,b_0^\dag)\right)e^{i\lambda^2n_c^2} \left(1+\frac{\gamma}{4\sqrt{2}} F_2(b_0,b_0^\dag)\right),
\end{split}
\end{equation}
where $F_{1(2)}(b_0,b_0^\dag)$ correspond to $f_{1(2)}$ after the switch. Summing up the two functions $F_1$ and $F_2$ and performing the partial trace on the mechanical oscillator initially in a thermal state, it is then possible to obtain at first order in $\gamma$ the effective unitary operator
\begin{equation}
\xi_{\sf eff}\simeq \exp\left\{i (\lambda^2 n_c^2  - \frac{\gamma}{2}(\lambda^4 n_c^4+3\lambda^2 n_c^2 \bar{n}) \right\}.
\end{equation}
In the limit $\lambda^2N_p^2\gg \bar{n}$, we get the mean value of the optical field shown in Eq.\eqref{afield} for an initial coherent state $|\alpha\rangle$
\begin{equation}
\begin{split}
 \langle a\rangle&=\langle\alpha|\xi_{\sf eff}^{\dag}a\xi_{\sf eff}|\alpha\rangle\\
 &=\alpha e^{-(|\alpha|^2+i\lambda^2)}\sum_{n=0}^\infty \frac{|\alpha|^{2n}}{n!}e^{-2i\lambda^2n}e^{-i\frac{\gamma}{2}\lambda^4(4n^3+6n^2+4n+1)}\\
&\simeq\alpha\langle a\rangle_0 e^{-i\frac{\gamma}{2}\lambda^4(4N_{p}^3+18N_p^2+10N_p+1)},
\end{split}
\end{equation}
where in the last step we have assumed $\gamma\lambda^4N_p^3\ll 1$.

For the sake of completeness we report here the exact result for the QFI
\begin{equation}\label{QFtot}\begin{split}
Q_\gamma &=4\left(\langle\psi'_{\gamma}|\psi'_{\gamma}\rangle-|\langle\psi'_{\gamma}|\psi_{\gamma}\rangle|^2\right)\\
  &\simeq \lambda^8\left(\langle\psi_{\gamma}|n_c^{8}|\psi_{\gamma}\rangle-\langle\psi_{\gamma}|n_c^4|\psi_{\gamma}\rangle^2\right)\\
  &\simeq \lambda^8 (16 N_p^7 + 216 N_p^6 + 964 N_p^5 + 1640 N_p^4 \\
  &\quad\; + 952 N_p^3 + 126 N_p^2 + N_p) \:.
\end{split}\end{equation}

\section{\label{appendix3}Cubic Anharmonicity}

In the case of a cubic anharmonicity the correction to the free Hamiltonian reads 
\begin{equation}\label{H'-anharmonicity}
\mathcal{H}_{an}=\frac{\delta}{3}\hbar\omega_m X_m^3 \:,
\end{equation}
where the parameter $\delta$ quantifies the anharmonicity. Again, following \cite{manko1982} we get the evolution for annihilation (creation) operator at the first order in $\delta$ and for initial displacements that satisfy $\delta\lambda N_p\ll 1$
\begin{equation}\label{b-time-3}\begin{split}
b(t) \simeq&\;b_0e^{-i\omega t}+\frac{\delta}{2^{3/2}}\bigg[\left(e^{-i\omega t}-1\right)\left(2b_0^\dag b_0+1\right)\\
&+\left(e^{-2i\omega t}-e^{-i\omega t}\right)b_0^2+(e^{-i\omega t}-e^{2i\omega t})\frac{b_0^{\dag 2}}{3}\bigg],
\end{split}\end{equation}
where in this case $\omega=\omega_m$ since the frequency is unperturbed at the first order in $\delta$. We highlight that we might exploit this feature to distinguish the two anharmonicities by looking at the revival in the visibility interference.
The overall evolution operator can thus be evaluated as in Eq.\eqref{Displacement4} by the anharmonic evolution of quadrature operators, which results in
\begin{equation}\label{Xmoretime-3}\begin{split}
  X_m(0)&=X_m\\
  X_m\left(\frac{\tau}{4}\right)&\simeq P_m+\delta(\Delta+b_0^{\dag 2}\nu+b_0^2\nu^*)\\
  X_m\left(\frac{\tau}{2}\right)&\simeq -X_m+\delta(2\Delta+\frac{1}{3}(b_0^{\dag 2}+b_0^2))\\
  X_m\left(\frac{3\tau}{4}\right)&\simeq -P_m+\delta(\Delta+b_0^{\dag 2}\nu^*+b_0^2\nu)
\end{split}\end{equation}
being $\Delta=-(b_0^\dag b_0+1/2)$ and $\nu=-(1/6)(2i+1)$. Going through the same procedure we showed in Appendix \ref{appendix1}, we recover the final effective evolution operator for the cavity field only (in the limit $\lambda^2N_p^2\gg \bar{n}$)
\begin{equation}\label{Displacement-final-3}\begin{split}
\xi_{\sf eff}\simeq \exp \{i ( \lambda^2 n_c^2 - \frac{2\delta}{9}\lambda^3 n_c^3 )\}.
\end{split}\end{equation}
From which we deduce that the optical field experiences a Kerr nonlinearity $\propto n_c^3$ entering into the cavity.
Hence, the mean value of the optical field after four pulses now results (in the limit $\delta\lambda^3N_p^2\ll 1$)
\begin{equation}\label{afield-3}\begin{split}
  \langle a\rangle=\langle\alpha |\xi_{\sf eff}^{\dag}a\xi_{\sf eff}|\alpha\rangle\simeq\alpha\langle a\rangle_0e^{-i\frac{2}{9}\delta\lambda^3 (3N_{p}^2+3N_p+1)}.
\end{split}\end{equation}

\noindent
As a result, the QFI for a cubic anharmonicity $\delta$, given an initial coherent state $|\alpha\rangle$ reads
\begin{equation}\label{QF}\begin{split}
  Q_\gamma&=4\left(\langle\psi'_{\gamma}|\psi'_{\gamma}\rangle-|\langle\psi'_{\gamma}|\psi_{\gamma}\rangle|^2\right)\\
  &\simeq \lambda^6\left(\langle\psi_{\gamma}|n_c^{6}|\psi_{\gamma}\rangle-\langle\psi_{\gamma}|n_c^3|\psi_{\gamma}\rangle^2\right)\\
  &\simeq \frac{16}{81} \lambda^6 (9 N_p^5 + 54 N_p^4 + 84 N_p^3 + 30 N_p^2 + N_p)
\end{split}\end{equation}
leading to the Cram\'er-Rao bound, 
\begin{equation}\label{CRB}\begin{split}
{\rm Var}(\delta)\geq\frac{1}{MQ_\gamma}\gtrsim \frac{9}{16 M \lambda^6N_{p}^5}
\end{split}\end{equation}

\section{\label{appendix3}Effect of losses}

In this appendix we evaluate the effect of losses on the unitary operator presented in Eq. \eqref{Displacement-final}. Since losses cause decreasing intensities for consecutive pulses, we can depict a lossy model through decreasing coupling strengths $\lambda_{i+1}/\lambda_{i}=1-\epsilon$ with $i=1,...,4$. 
By following the same procedure as in Appendix A, the evolution operator reads 
\begin{equation}\label{U-losses}\begin{split}
 &U\simeq e^{-i\lambda_4n_cP_m}e^{\frac{\gamma}{4}f_1(b_0,\bd_0)}e^{i\lambda_4n_cP_m}\\
 &\quad\quad\times\xi_{h}e^{-i\lambda_1n_cX_m}e^{\frac{\gamma}{4}f_2(b_0,\bd_0)}e^{i\lambda_1n_cX_m}
\end{split}\end{equation}
where $\xi_{h}$ is the harmonic displacement given by
\begin{equation}\label{harm-losses}
 \xi_{h}=D(n_c\mu)e^{in^2_c[\lambda_3\lambda_2+\frac{1}{2}(\lambda_2-\lambda_4)(\lambda_1-\lambda_3)]}.
\end{equation}
$D(n_c\mu)=e^{n_c(\mu\bd-\mu^*b)}$ with $\mu=(1/\sqrt{2})[(\lambda_4-\lambda_2)+i(\lambda_1-\lambda_3)]$ is a displacement operator that does not allow light and mirror to be disentangled after a closed loop. The functions $f_1(b_0,\bd_0)$ and $f_2(b_0,\bd_0)$ have the same formal definitions as in \eqref{f}  with $\lambda\rightarrow\lambda_4$ and $\lambda\rightarrow\lambda_1$, respectively. 
Calculating the exponentials in \eqref{U-losses} and expanding at the first order in $\gamma$ we get
\begin{equation}\label{U-losses-2}\begin{split}
 U\simeq  \xi_{h}+\frac{\gamma}{4}[F_1(b_0,\bd_0) \xi_{h}+ \xi_{h}F_2(b_0,\bd_0)]
\end{split}\end{equation}
In order to estimate the contribution of losses, we perform the partial trace over the mechanical degrees of freedom
\begin{equation}\label{harm-losses}\begin{split}
 &\langle \xi_{h}\rangle=e^{-\frac{|\mu|^2}{2}n_c^2(1+2\bar{n})}e^{in^2_c[\lambda_3\lambda_2+\frac{1}{2}(\lambda_2-\lambda_4)(\lambda_1-\lambda_3)]}\\
 &\frac{\gamma}{4}\langle F_1(b_0,\bd_0) \xi_{h}+ \xi_{h}F_2(b_0,\bd_0)\rangle\simeq\frac{\gamma}{4}\lambda^4n_c^4+O(\epsilon\lambda^4n_c^3\bar{n}).
\end{split}\end{equation}
It is therefore possible to neglect the effect of losses on the anharmonic contribution to the unitary operator in the limit $\epsilon\bar{n}\ll N_p$. We also point out that losses change also the harmonic term (and phase) giving rise to a reduction of visibility that can be estimated before performing the experiment.


\begin{references}
\bibitem{aspelmeyer2014} M. Aspelmeyer, T. Kippenberg and F. Marquardt, Rev. Mod. Phys. {\bf 86}, 1391 (2014).
\bibitem{arcizet2006} O. Arcizet, P. F. Cohadon, T. Briant, M. Pinard, A. Heidmann, Nature {\bf 444}, 71 (2006).
\bibitem{gigan2006} S. Gigan et al., Nature 444, 67 (2006).
\bibitem{thompson2008} J. D. Thompson, B. M. Zwickl, A. M. Jayich, F. Marquardt, S. M. Girvin, and J. G. E. Harris, Nature {\bf 452}, 72 (2008).
\bibitem{barker2010} P. F. Barker and M. N. Shneider, Phys. Rev. A {\bf 81} 023826 (2010).
\bibitem{chang2010} D. E. Chang et al, Proc. Natl Acad. Sci. USA {\bf 107}, 1005 (2010).
\bibitem{pflanzer2012} A. C. Pflanzer, O. Romero-Isart, and J. I. Cirac, Phys. Rev. A {\bf 86} 013802 (2012).

\bibitem{caves1980} C. M. Caves, K. S. Thorne, R. W. P. Drever, V. D. Sandberg, and M. Zimmermann, Rev. Mod. Phys. {\bf 52}, 341 (1980).
\bibitem{braginsky1995} V.B. Braginsky and F.Y.A. Khalili, \textit{Quantum measurements} (Cambridge, 1995).
\bibitem{romeroisart2011} O. Romero-Isart, Phys. Rev. A {\bf 84}, 052121 (2011).
\bibitem{bahrami2014} M. Bahrami, M. Paternostro, A. Bassi, and H. Ulbricht, Phys. Rev. Lett. \textbf{112} 210404 (2014).
\bibitem{pikovski2012} I. Pikovski,	M.R. Vanner,	M. Aspelmeyer, M.S. Kim and C. Brukner, Nature Physics \textbf{8}, 393 (2012).
\bibitem{bawaj2015} M. Bawaj, C. Biancofiore, M. Bonaldi, F. Bonfigli, A. Borrielli, G. Di Giuseppe, L. Marconi, F. Marino, R. Natali, A. Pontin, G.A. Prodi,	E. Serra, D. Vitali, and F. Marin, Nat. Commun. \textbf{6}, 7503 (2015).



\bibitem{kronwald2013} A. Kronwald, F. Marquardt and A. A. Clerk, Phys. Rev. A {\bf 88} 063833 (2013).
\bibitem{genoni2015a} M. G. Genoni, M. Bina, S. Olivares, G. De Chiara and M. Paternostro, New J. Phys. {\bf 17}, 013034 (2015).
\bibitem{genoni2015b} M. G. Genoni, J. Zhang, J. Millen, P. F. Barker and A. Serafini, New J. Phys. {\bf 17} 073019 (2015).
\bibitem{wollman2015} E. E. Wollman et al., Science {\bf 349}, 952 (2015).

\bibitem{pirkkalainen2015} J.-M. Pirkkalainen, E. Damsk\"agg, M. Brandt, F. Massel, and M. A. Sillanp\"a\"a, Phys. Rev. Lett. {\bf 115}, 243601 (2015).
\bibitem{rips2012} S. Rips, M. Kiffner, I. Wilson-Rae, and M. J. Hartmann, New J. Phys. {\bf 14} 023042 (2012).

\bibitem{qian2012} J. Qian, A. A. Clerk, K. Hammerer, and F. Marquardt, Phys. Rev. Lett. {\bf 109}, 253601 (2012).
\bibitem{borkje2014} K. Borkje, Phys. Rev. A {\bf 90}, 023806 (2014).

\bibitem{bose1997} S. Bose, K. Jacobs, and P.L. Knight, Phys. Rev. A \textbf{56}, 4175, (1997).
\bibitem{penrose2003} W. Marshall, C. Simon, R. Penrose, and D. Bouwmeester, Phys. Rev. Lett \textbf{91}, 130401 (2003).
\bibitem{lombardo2015} D. Lombardo and J. Twampley, Sci. Rep. {\bf 5} 13884 (2015).
\bibitem{Dykman2012} M. Dykman, \textit{Fluctuating nonlinear oscillators: from nanomechanics to quantum superconducting circuits}, Oxford University Press (2012).
\bibitem{Eichler2011} A. Eichler, J. Moser, J. Chaste, M. Zdrojek, I. Wilson-Rae and A. Bachtold, Nature Nanotech. {\bf 6}, 339 (2011).



\bibitem{gieseler2013} J. Gieseler, L. Novotny and R. Quidant, Nature Physics {\bf 9}, 806 (2013).
\bibitem{rips2014} S. Rips, I. Wilson-Rae and M. J. Hartmann, Phys. Rev. A {\bf 89} 013854 (2014).
\bibitem{fonseca2015} P. Z. G. Fonseca, E. B. Aranas, J. Millen, T. S. Monteiro andP. F. Barker, arXiv:1511.08482 (2015).
\bibitem{milburn1986} G.J. Milburn and C.A. Holmes, Phys. Rev. Lett. \textbf{56}, 2237 (1986).
\bibitem{joshi2011} C. Joshi, M. Jonson, E. Andersson and P. {\"O}hberg, J. Phys. B \textbf{44}, (24), 245503 (2011).
\bibitem{lu2015} X.-Y. L\''u, J.-Q. Liao, L. Tian, and F. Nori, Phys. Rev. A {\bf 91}, 013834 (2015).
\bibitem{teklu2015} B. Teklu, A. Ferraro, M. Paternostro and M.G.A. Paris,  arXiv:1501.03767v1 (2015).
\bibitem{paris2014} M.G.A. Paris, M. G. Genoni, N. Shammah and B. Teklu, Phys. Rev. A {\bf 90}, 012104 (2014).
\bibitem{vanner2011} M.R. Vanner, I. Pikovski, G.D. Cole, M.S. Kim, C. Brukner, K. Hammerer, G.J. Milburn, and M. Aspelmeyer, Proc. Natl Acad. Sci. USA \textbf{108}, 16182 (2011).
\bibitem{khosla2013} K. E. Khosla, M. R. Vanner, W. P. Bowen, and G. J. Milburn, New J. Phys. {\bf 15} 043025 (2013).
\bibitem{law1995} C.K. Law, Phys. Rev. A \textbf{51}, 2537 (1995).
\bibitem{landau1976} L.D. Landau and E.M. Lifshitz, \textit{Mechanics} (Butterworth-Heinemann, 1976).
\bibitem{manko1982} M.G. Krivoshlykov, V.I. Mank'ko and I.N. Sissakian, Phys. Lett. \textbf{90 A}, 165 (1982).
\bibitem{mancini1997} S. Mancini, V.I. Man'ko, and P. Tombesi,, Phys. Rev. A \textbf{55}, 3042, (1997).

\bibitem{higgins2007} B. L. Higgins et al., Nature {\bf 450}, 393 (2007).
\bibitem{brivio2010} D. Brivio, S. Cialdi, S. Vezzoli, B. T. Gebrehiwot, M. G. Genoni, S. Olivares, and M. G. A. Paris, Phys. Rev. A 81, 012305 (2010).
\bibitem{berni2015} A. A. Berni,	T. Gehring, B. M. Nielsen, V. H\"andchen, M. G. A. Paris, and U. L. Andersen, Nature Photonics {\bf 9}, 577 (2015).
\bibitem{armata-latmiral} F. Armata, L. Latmiral, I. Pikovski, M. R. Vanner, \v{C}. Brukner, and M. S. Kim, arXiv: 1604.05679.
\bibitem{aldana2013} S. Aldana, C. Bruder and A. Nunnenkamp, Phys. Rev. A, \textbf{88} 043826 (2013).
\bibitem{paris2009} M.G.A. Paris, Int. J. Quant. Inf. 7, 125 (2009).
\bibitem{ferraro2005} A. Ferraro, S. Olivares and M.G.A. Paris, \textit{Gaussian States in Quantum Information}, (Bibliopolis, Napoli, 2005).
\bibitem{genoni2014} M.G. Genoni, S. Mancini and A. Serafini, Russian Journal of Mathematical Physics 21, 329 (2014).
\bibitem{suzuki1977} M. Suzuki, Comm. Math. Phys. \textbf{57}, 193-200 (1977).
\end{references}
\end{document}